\documentclass[aps,pre,twocolumn,superscriptaddress,amsfonts,showpacs,floatfix]{revtex4}
\usepackage{epsfig}
\usepackage {array}
\usepackage {amsmath}
\usepackage {lscape}
\usepackage{graphicx}
\usepackage[dvips]{color}

\newcommand{\Pm}{\mathrm{Pm}}

\providecommand\bB{\boldsymbol{\rm B}}

\providecommand\bF{\boldsymbol{\rm F}}

\providecommand\bOmega{\boldsymbol{\Omega}}
\providecommand\bu{\boldsymbol{\rm u}}
\providecommand\bv{\boldsymbol{\rm v}}

\begin{document}
\title{Cascades and dissipation ratio in rotating MHD turbulence at low magnetic Prandtl number.}
\author{Franck Plunian}
\affiliation{Universit\'e Joseph Fourier, CNRS / INSU, Laboratoire de G\'eophysique
Interne et Tectonophysique, B.P. 53, 38041 Grenoble Cedex 9, France}
\author{Rodion Stepanov}
\affiliation{Institute of Continuous Media Mechanics, Korolyov 1, 614013 Perm, Russia}

\begin{abstract}
A phenomenology of isotropic magnetohydrodynamic turbulence subject to both rotation and applied magnetic field is presented. It is assumed that the triple correlations decay-time is the shortest between the eddy turn-over time and the
ones associated to the rotating frequency and Alfv\'en wave period. For $\Pm=1$ it leads to four kinds of piecewise spectra,
depending on the four parameters, injection rate of energy, magnetic diffusivity, rotation rate and applied field.
With a shell model of MHD turbulence (including rotation and applied magnetic field), spectra for $\Pm \le 1$ are presented,
together with the ratio between magnetic and viscous dissipation.
\today
\end{abstract}
\pacs{47.27.E-, 47.65.-d, 96.50.Tf}
\maketitle
\section{Introduction}
Magnetohydrodnamic turbulence in natural objects is often subject to global rotation or applied magnetic field, or both.
In the Earth's core the turbulence occurs under the fast rotation of the planet and is embedded in the dipolar magnetic field
produced by dynamo action. Such double effect is currently studied in an experiment with liquid sodium \cite{Schmitt08}.
Waves of different types have been measured that might be attributed to either Alfv\'en or Rossby waves or a combination of both.
The frequency spectra show a series of bumps, attributed to wave frequencies, in addition to piecewise slopes.
A proper understanding of such rotating MHD-turbulence would require a non-isotropic formalism.
 Several ones
have been developed for fast rotation \cite{Galtier03,Schaeffer06,Bellet06}. Phenomenological approaches relying on three-wave \cite{Galtier05} or four-wave \cite{Goldreich97} resonant interactions have been developed for an applied field and documented numerically \cite{Bigot08}.

In the present paper we come back to the Iroshnikov \cite{Iroshnikov63} and Kraichnan \cite{Kraichnan65} phenomenology for isotropic MHD turbulence. They argue that the destruction of phase coherence by Alfv\'en waves traveling in opposite directions introduces a new time-scale $\tau_A$. It might control the energy transfer, provided it is shorter than the eddy turn-over time-scale $\tau_K$.
Applying the same idea, Zhou \cite{Zhou95} suggests that due to global rotation the kinetic energy spectrum is affected through phase scrambling, leading to a third time-scale $\tau_{\Omega}$ associated to the rotation frequency. The generalization to both global rotation and applied magnetic field is therefore straightforward (see section \ref{Phenomenology}), the energy transfers being controlled by the shortest time-scale between $\tau_K$, $\tau_A$ and $\tau_{\Omega}$.

An advantage of assuming isotropy is that it can be tested against simulations with shell models.
Shell models are toy-models that mimic the original Navier-Stokes and induction equations projected in Fourier space, within shells which are logarithmically spaced.
There are only two complex variables per shell, one corresponding to the velocity, the other to the magnetic field
\cite{Frick98,Stepanov06}.
Depending on the model, the energy transfers can be considered as local or not \cite{Plunian07}. Such models
allow for simulations at realistically low viscosity $\nu$ and magnetic Prandtl number $\Pm=\nu/\eta$ \cite{Stepanov08}, where $\eta$ is the magnetic diffusivity.
The time dependency of the solutions is strongly chaotic, eventually leading to intermittency. Therefore, though all geometrical details of velocity and magnetic fields are lost, shell models give relevant informations on spectral quantities like energies, helicities, energy transfers, etc. In section \ref{shell model} we introduce such a shell model of rotating MHD turbulence, taking care to keep the terms corresponding to rotation and applied magnetic field as simple as possible. For $\Pm\le 1$ we calculate the spectra for different values of rotation $\Omega$ and applied field $V_A$.
We also calculate the ratio of the joule dissipation over the viscous dissipation, which cannot be estimated from scaling laws.

\section{Phenomenology}
\label{Phenomenology}
\subsection{Time scales}
Following \cite{Kraichnan65} (see also \cite{Zhou95} and \cite{Matthaeus89}), we assume that for
homogeneous isotropic statistically steady turbulence the decay of triple correlations,
occurring in a time scale $\tau_3(k)$, is responsible for the turbulent spectral transfer
$\varepsilon$ from wavenumbers lower than $k$ to higher wavenumbers.
This implies $\tau_3(k) \sim \varepsilon$. Assuming in addition that $\varepsilon$
depends only on the wave number $k$ and the kinetic energy spectral density $E(k)$,
a simple dimensional analysis
leads to
\begin{equation}
	\varepsilon \sim \tau_3(k) E^2(k) k^4.
\end{equation}
The kinetic energy spectral density is defined as $E(k)=k^{-1}u^2(k)$ where $u(k)$ is the
characteristic velocity of eddies at scale $k$.\\
In absence of applied magnetic field and rotation, the time scale for $\tau_3(k)$ is the
eddy turn-over time
\begin{equation}
	\tau_{K}(k)=\left[ku(k)\right]^{-1},
\end{equation}
leading to the Kolmogorov turbulence energy spectrum $E(k) \sim \varepsilon^{2/3}
k^{-5/3}$.\\

For fully developed MHD turbulence at $\Pm=1$ the same Kolmogorov spectrum
is assumed for both kinetic and magnetic energy provided that the system is much above the onset for dynamo action \cite{Stepanov06}. In that case
$E(k)$ denotes either the kinetic or magnetic energy spectral density.
In presence of an applied magnetic field $\bB_0$ an other
possible time scale for $\tau_3(k)$ is the Alfv\'en time scale
\begin{equation}
	\tau_A (k)=\left[kV_A\right]^{-1},
\end{equation}
leading to the Alfv\'en turbulence energy spectrum
$E(k) \sim V_A^{1/2} \varepsilon^{1/2} k^{-3/2}$.\\
Finally for rotating turbulence caused by uniform rotation $\Omega$ a third possible
time scale for $\tau_3(k)$ is the rotating frequency
\begin{equation}
	\tau_{\Omega}=\Omega^{-1},
\end{equation}
leading to the rotating turbulence energy spectrum
$	E(k) \sim \Omega^{1/2} \varepsilon^{1/2} k^{-2}$.

The value of $\tau_3(k)$ is naturally defined by
\begin{equation}
	\tau_3(k)=\min\left\{\tau_{K}(k),\tau_A (k), \tau_{\Omega} \right\}.
\end{equation}
It corresponds to the fastest way to transfer energy to smaller scales, between non-linear eddy cascade,
Alf\'en waves interactions and phase scrambling due to rotation.
In addition we define the magnetic dissipation time scale by
\begin{equation}
	\tau_{\eta}(k)=(k^2 \eta)^{-1}.
\end{equation}
The dissipation range corresponds to $k\ge k_{\eta}$
with $k_{\eta}$ defined
by $\tau_3(k_{\eta})=\tau_{\eta}(k_{\eta})$.

Therefore at each scale $k^{-1}$, we
have to compare the four time scales $\tau_{K}(k),\tau_A (k), \tau_{\Omega}$ and
$\tau_{\eta}(k)$ to figure out what kind of turbulence occurs.

\subsection{Spectra for $\Pm=1$}
At $k\ll1$, $\tau_{\Omega} < \min\left\{\tau_{K}(k), \tau_A (k), \tau_{\eta} (k) \right\}$
implying that $\tau_3(k)=\tau_{\Omega}$, unless $\Omega=0$.
This corresponds to a rotating turbulence with $E(k)=\Omega^{1/2}\varepsilon^{1/2}
k^{-2}$.
For larger $k$, $\tau_K(k), \tau_A(k)$ and $\tau_{\eta}(k)$ decrease while $\tau_{\Omega}$
stays constant. Therefore, provided that the dissipation is not too strong, a first
transition occurs at a scale for which
$\tau_{\Omega}=\min\left\{\tau_K(k),\tau_A(k)\right\}$. This scale is, either
(i) $k_1=(\Omega^3/\varepsilon)^{1/2}$ if $\varepsilon \ge \Omega V_A^2$, or
(ii) $k_1=\Omega/V_A$ if $\varepsilon \le \Omega V_A^2$.
This transition leads to, either (i) a Kolmogorov $E(k)=\varepsilon^{2/3}k^{-5/3}$, or
(ii) an Alfv\'en $E(k)=V_A^{1/2}\varepsilon^{1/2}k^{-3/2}$ turbulence. This transition
does not occur if the dissipation overcomes the Kolmogorov and Alfv\'en turbulence, namely
if (i) $\eta \ge \varepsilon / \Omega^2$ and (ii) $\eta \ge V_A^2 / \Omega$. In
that case the dissipation scale is given by $k_{\eta}=(\Omega/\eta)^{1/2}$.

In case (i) provided again that dissipation is not too strong, a second transition occurs
at $k_2=\varepsilon / V_A^3$ . This transition leads to an Alfv\'en turbulence
$E(k)=V_A^{1/2}\varepsilon^{1/2}k^{-3/2}$ until the dissipation becomes dominant for
$k\ge k_{\eta}$ with $k_{\eta}=V_A/\eta$.
If $V_A^4\le \eta \varepsilon$ the dissipation overcomes the Alfv\'en turbulence and
the
dissipation scale is given by $k_{\eta}=\varepsilon^{1/4}\eta^{-3/4}$.

In case (ii) a second transition toward a Kolmogorov turbulence is not possible. Indeed,
it would occur at $k=\varepsilon / V_A^3$ which can not be larger than $k_1$ from the
condition $\varepsilon \le \Omega V_A^2$. In that case the Alfv\'en turbulence simply
extends to the dissipation scale given by $k_{\eta}=V_A/\eta$.

\begin{figure*}
\begin{tabular}{@{\hspace{0cm}}c@{\hspace{-7.3cm}}c@{\hspace{1cm}}c@{\hspace{5.7cm}}c@{}}
 \includegraphics[width=0.6\textwidth]{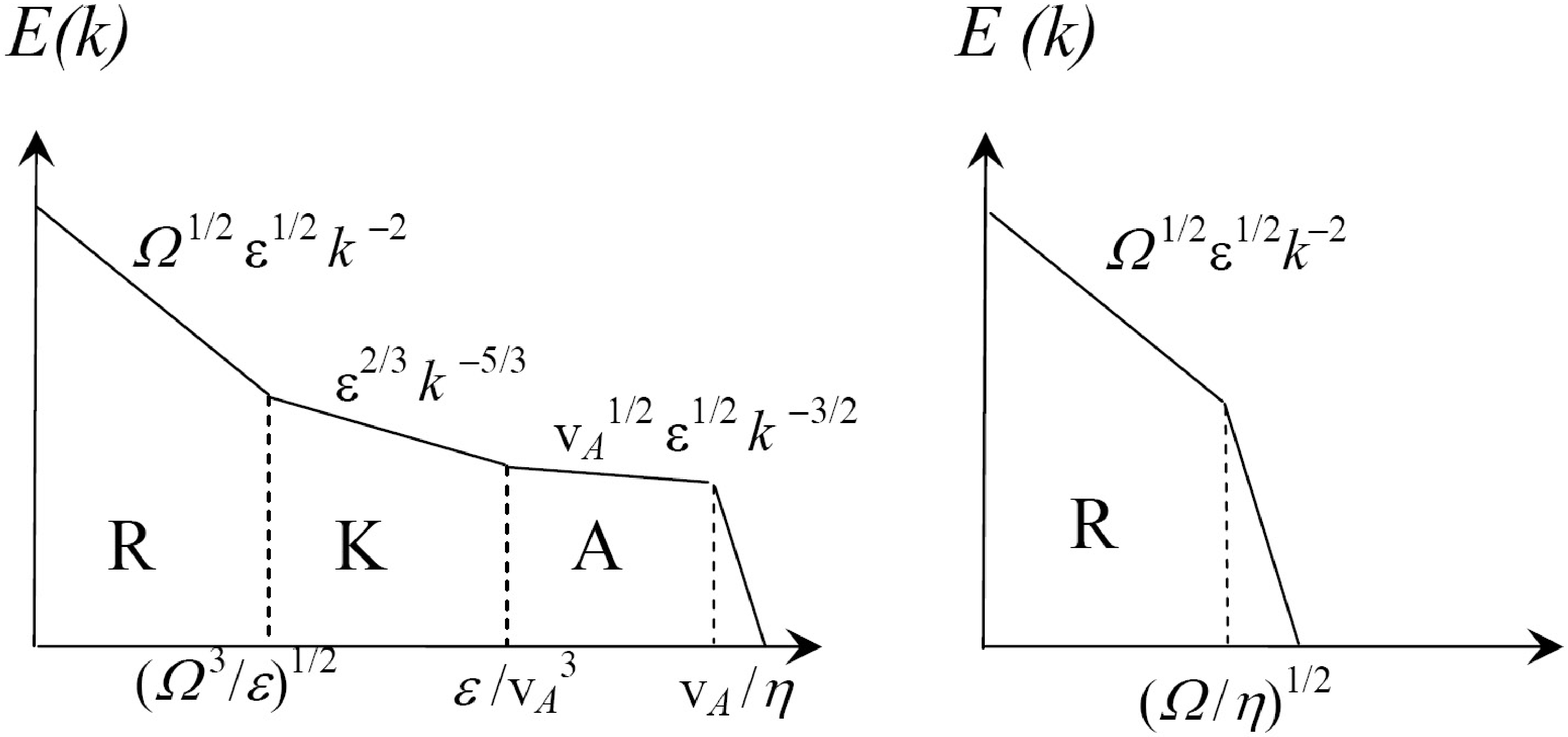}&&\raisebox{0.8cm}{\hspace{-0.5cm}$k$}&\raisebox{0.8cm}{\quad\quad$k$}
 \\*[0cm]
 \includegraphics[width=0.6\textwidth]{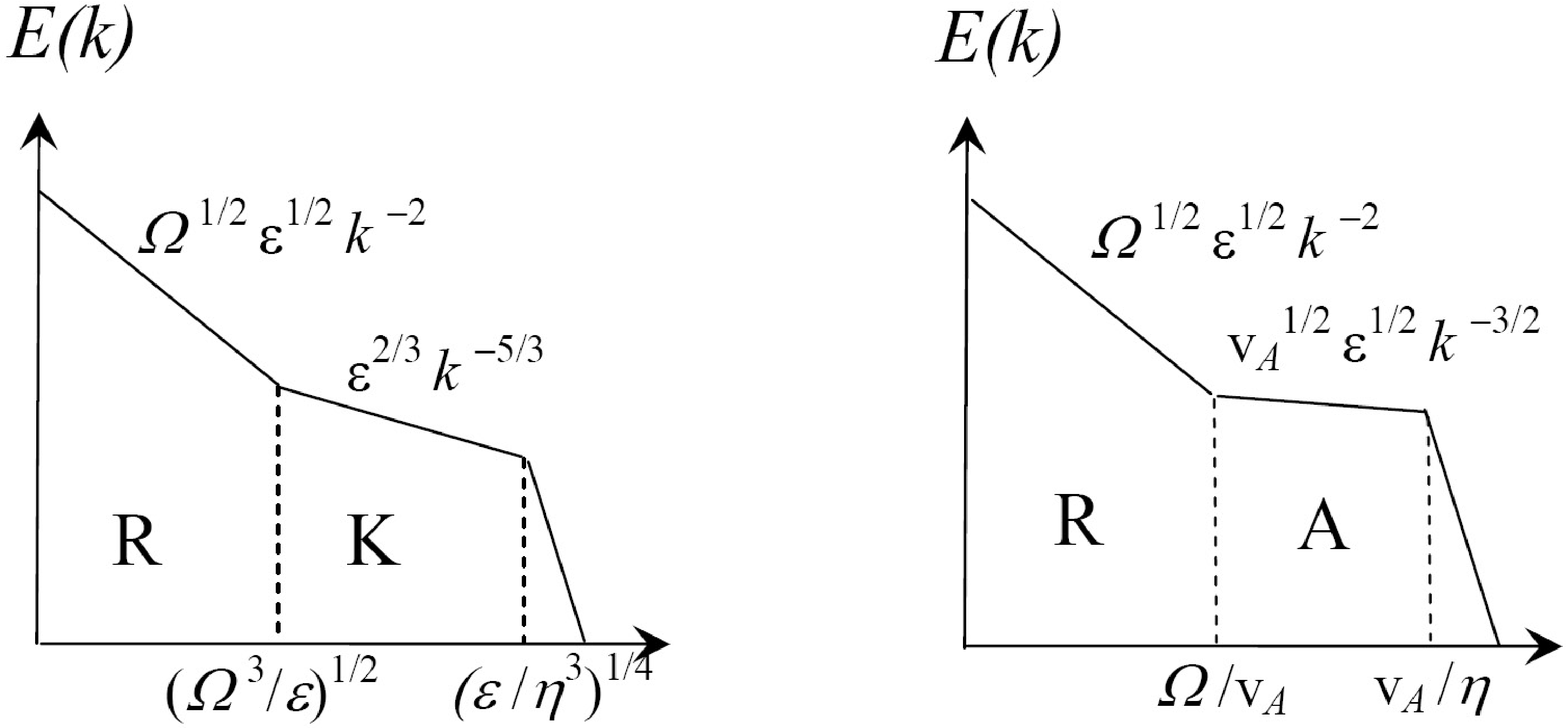}&\raisebox{0.8cm}{\hspace{-1.cm}$k$}&&\raisebox{0.8cm}{\quad\quad$k$}\\*[0cm]
 \end{tabular}
\caption{Possible inertial regimes of energy spectral density in rotating MHD turbulence for $\Pm=1$. The capital letters $R, A$ and $K$ denote a
rotating, Alfv\'en or Kolmogorov turbulence.}
\label{Spectra}
\end{figure*}

\begin{figure}
\includegraphics[width=0.45\textwidth]{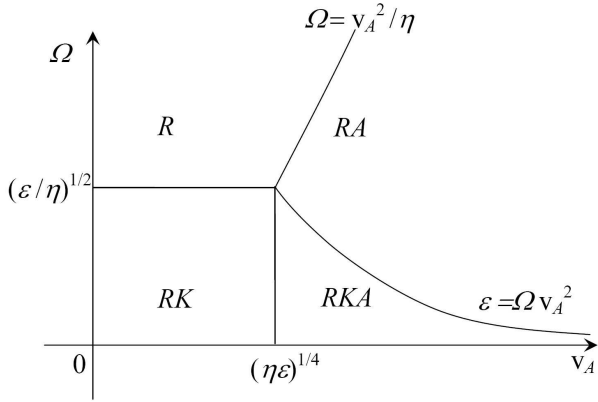}
\caption{The four possible turbulent inertial regimes given in the map $(V_A,\Omega)$.}
\label{map}
\end{figure}

The four possible types of inertial regimes are sketched in Fig.~\ref{Spectra} in which the spectral energy density is
plotted versus $k$ for $\Pm=1$. The slopes and characteristic wave numbers are indicated.
The
conditions to get one of these four possible inertial regimes are summarized in the plane
$(V_A, \Omega)$ in Fig.~\ref{map}.
The case without rotation corresponds to the abscissa axis. Then two regimes KA
or K are possible depending whether $\eta / V_A^2 \le V_A^2 / \varepsilon$ or not. The
case without applied magnetic field corresponds to the vertical axis.
Then the two regimes R or RK are possible depending whether $\eta \Omega/\varepsilon
\ge \Omega^{-1}$ or not. Without both rotation and applied magnetic field
a K type of turbulence is found.

From our analysis we note that inertial regimes of type AK or RAK are never possible. On the other hand
inertial regimes of type KA, A, or K are possible provided the forcing scale is sufficiently small.
In Fig.~\ref{Spectra} it corresponds to begin the spectra at a larger wave number.
For $\Pm<1$ the inertial range of the kinetic energy spectrum prolongates at scales smaller than $k_{\eta}$
with either an R, K or RK spectrum.

\section{Shell model}
\label{shell model}
\subsection{The model}
The equations of MHD turbulence for an incompressible fluid embedded in an external uniform
magnetic field $\bB_0$ and subject to rotation $\bOmega$ write
\begin{eqnarray}
\partial \bu/\partial t &+& (\bu \cdot \nabla)\bu - ((\bB+\bv_A) \cdot \nabla)\bB + 2\bOmega
\times \bu \nonumber
	\\&=& \nu \nabla^2 \bu + \bF - \nabla P_t \label{U}\\
	\partial \bB/\partial t &+& (\bu \cdot \nabla)\bB - ((\bB+\bv_A) \cdot \nabla)\bu = \eta
\nabla^2 \bB \label{B}\\
	\nabla \cdot \bu &=& \nabla \cdot \bB = 0 \label{incomp}
\end{eqnarray}
in which $\textbf{v}_A=\bB_0 / \sqrt{\mu \rho}$ is the Alfv\'en velocity (where
$\mu$ and $\rho$ are respectively the fluid magnetic permeability and density) and $\bB$ is given in unit of $V_A=|\bv_A|$. The total
pressure $P_t = P + b^2 /2$ is a functional of $\bu$ and $\bB$ owing to the
incompressibility condition (\ref{incomp}). The forcing $\bF$ insures the fluid motion.

From (\ref{U}) (\ref{B}) (\ref{incomp}) we derive the following shell model
\begin{eqnarray}
\dot{U}_n &=& i k_n \left[Q_n(U,U)-Q_n(B,B)\right] \nonumber \\
&+& i k_n V_A(t) B_n + i \Omega(t)
U_n
- \nu k_n^2 U_n + F_n(t), \label{eq_u} \\
\dot{B}_n &=& i k_n \left[Q_n(U,B)-Q_n(B,U)\right] \nonumber \\&+& i k_n V_A(t) U_n
- \eta k_n^2 B_n, \label{eq_b}
\end{eqnarray}
where
\begin{eqnarray}
Q_n(X,Y)= \lambda^2 (X_{n+1}Y_{n+1}+X_{n+1}^*Y_{n+1}^*)
-X_{n-1}^r Y_n \nonumber \\ -X_n Y_{n-1}^r+{\rm i} \lambda(2
X_n^*Y_{n-1}^i+X_{n+1}^r Y_{n+1}^i-X_{n+1}^i Y_{n+1}^r) \nonumber
\\
\label{Qn}
+X_{n-1}Y_{n-1}+X_{n-1}^*Y_{n-1}^* -\lambda^2(X_{n+1}^r Y_n
 +X_n Y_{n+1}^r) \nonumber \\+{\rm i} \lambda(2
X_n^*Y_{n+1}^i+X_{n-1}^r Y_{n-1}^i-X_{n-1}^i Y_{n-1}^r), \label{shellnew}
\end{eqnarray}
represents the non linear transfer rates and $F_n$ the turbulence forcing. This model
is based on wavelet decompostion \cite{Zimin95}. Compared to other shell models
\cite{Gledzer73,Ohkitani89,Lvov98} it has the advantage that helicities are much better defined,
like those based on helical wave decomposition \cite{Benzi96a,Lessinnes09a,Lessinnes09b}.
It
has been introduced in
its hydrodynamic form to study spectral properties of helical turbulence \cite{Stepanov09},
and in its MHD form to study cross-helicity effect on cascades \cite{Mizeva09}.
The parameter $\lambda$ is the geometrical factor from which the wave number is defined $k_n=k_0 \lambda^n$.
As explained in \cite{Plunian07} an optimum shell spacing is the golden number
$\lambda = (1 + \sqrt{5}) / 2$.
The terms involving $\Omega$ and $V_A$ were already introduced in several previous papers dealing with either rotation \cite{Hattori04,Chakraborty10} or applied magnetic field \cite{Biskamp94,Hattori01}.

\subsection{Conservative quantities}
Expressions for the kinetic energy and helicity, $E_U$ and $H_U$, magnetic energy and
helicity, $E_B$ and $H_B$, and cross helicity $H_C$, are given by
\begin{eqnarray}
E_U &=& \sum_{n}E_U(n), \; E_U(n)=\frac{1}{2} |U_n|^2, \\
H_U &=& \sum_{n}H_U(n), \; H_U(n)=\frac{i}{2} k_n ((U_n^*)^2-U_n^2), \\
E_B &=& \sum_{n}E_B(n), \; E_B(n)=\frac{1}{2} |B_n|^2, \label{kinetic}\\
H_B &=& \sum_{n}H_B(n), \; H_B(n)=\frac{i}{2} k_n^{-1}((B_n^*)^2-B_n^2),
\label{magnetic} \\
H_C &=& \sum_{n}H_C(n), \; H_C(n)=\frac{1}{2}  (U_n B_n^* + B_n U_n^*).\label{crosshelicity}
\end{eqnarray}

In the inviscid and non-resistive limit ($\nu=\eta=0$), the total energy
$E=E_U+E_B$, magnetic helicity and cross helicity must be conserved
($\dot{E}=\dot{H}_B=\dot{H}_C=0$).
Here with the additional Coriolis and Alfv\'enic terms the properties of conservation are not necessarily satisfied.
A summary of theses properties is given in table \ref{conservation} for 3D MHD turbulence.
In the case of pure hydrodynamic turbulence (without magnetic field) the kinetic energy
and helicity must be conserved ($\dot{E_U}=\dot{H}_U=0$) even with Coriolis forces.
\begin{table}
\begin{center}
\begin{tabular}{|@{\hspace{0.cm}}c|@{\hspace{0.2cm}}l|@{\hspace{0.2cm}}l|@{\hspace{0.2cm}}l|@{\hspace{0.2cm}}l|@{}}
$\Omega$	&	=0	&	$\neq0$	&	$=0$			&	$\neq0$					\\*[0cm]
$V_A$       & =0  & $=0$		&	$\neq 0$	&	$\neq0 $				\\*[0cm]
\hline
$E$     		&	Y	  &	Y       &	Y         &	Y               \\*[0cm]
$H_C$   		& Y   &	N       &	Y         &	N               \\*[0cm]
$H_M$   		& Y   &	Y       &	N         &	N               \\*[0cm]
	\end{tabular}	
	\caption{In 3D MHD turbulence, conservation properties of total energy $E$, cross-helicity $H_C$ and magnetic helicity $H_M$
	depending on global rotation $\Omega$ and applied field $V_A$.}
	\label{conservation}
	\end{center}
\end{table}
\subsection{Time-scales}
In (\ref{eq_u}) and (\ref{eq_b}) the forcing $F_{n_F}(t)$ (applied at some scale $k_{n_F}^{-1}$), the global rotation $\Omega(t)$ and the applied field
$V_A(t)$ have constant intensities $|F_{n_F}|, \Omega$ and $V_A$. Only their sign may change after a period of time
$t_F$, $t_{\Omega}$ and $t_{V_A}$, the probability of changing from one period to the next being random.
Such a trick allows to control the two characteristic times $\tau_{\Omega} \approx t_{\Omega}$ and
$\tau_{V_A} \approx t_{V_A}$. In the simulations we take $t_{\Omega}=1/\Omega$ and $t_{V_A}=1/(k_{n_F}V_A)$.
It is in same spirit than the one used in \cite{Hattori01} and \cite{Hattori04} though much simpler.
Incidentally the random change of sign of $\Omega(t)$ insures that there is no injection of kinetic helicity on average.
Taking a random sign in $F_{n_F}(t)$ we insure that the forcing intensity satisfies $|F_{n_F}|\approx \sqrt{2 \varepsilon / t_F}$.
It is also important that $t_F$ is the shortest among all other characteristic times of the problem $\tau_K, \tau_{\Omega}$ and $\tau_{V_A}$ (and of course $\tau_{\eta}$). We choose  $t_F\le \frac{1}{10}\min\left\{\tau_K, \tau_{\Omega}, \tau_{V_A}\right\}$.

\subsection{No injection of cross-helicity}
In addition it is important to control the injection of cross-helicity as was shown in \cite{Mizeva09}. Indeed any spurious injection
of cross-helicity may lead to a supercorrelation state where $U_n \approx B_n$ implying equality not only in intensity (as in equipartition) but also in phase. In that case the flux of kinetic energy is depleted, implying an accumulation of energy at large scale and steeper spectral slopes. In order to compare the results to the phenomenological approach we impose the injection of cross-helicity to be zero.
For that we could use the forcing
\begin{equation}
\frac{F_{n_F}}{|F_{n_F}|}=\pm i \frac{B_{n_F}}{|B_{n_F}|}
\label{forcing1}
\end{equation}
where again the
sign is randomly changed after each period of time $t_F$.
This forcing is however ill-defined as soon as $|B_{n_F}|\ll |U_{n_F}|\approx 1$. To fix this problem we use the following forcing
\begin{equation}
\frac{F_{n_F}}{|F_{n_F}|}=\frac{a e^{i \varphi} \pm i \zeta \frac{B_{n_F}}{|B_{n_F}|}}{a + \zeta}
\end{equation}
with $\zeta=|B^2_{n_F}|/|U^2_{n_F}|$,
in which $\varphi$ is a phase randomly changed after each period of time $t_F$, and $a$ an additional parameter.
In the case $\zeta \gg a$, (\ref{forcing1}) is recovered, and the phase of $F_{n_F}$ is mainly determined by the phase of $B_{n_F}$ so that it corresponds to zero injection of cross-helicity.
In the case $\zeta \ll a$   the phase of $F_{n_F}$ is controlled by the random phase $\varphi$. Since $B_{n_F}$ is small there is no cross-helicity injection too.
The value $a=10^{-6}$ provides a robust forcing with always a low level of cross-helicity.

\subsection{Dissipations}
We define the dissipation of $U$ and $B$ at scale $k_n$ by $D_U(k_n)=\nu k_n^2 |U_n|^2$ and
$D_B(k_n)=\eta k_n^2 |B_n|^2$.
From the phenomenological formalism above we expect the total dissipation to be equal to the injection rate of energy at the forcing scale
$\varepsilon_{\nu}+\varepsilon_{\eta}=\varepsilon$, with $\varepsilon_{\nu}=\sum_n D_U(k_n)$ and $\varepsilon_{\eta}=\sum_n D_B(k_n)$.
Equivalently in pure HD we would have $\varepsilon_{\nu}=\varepsilon$.
On the other hand the ratio of both dissipations $\rho=\varepsilon_{\eta}/\varepsilon_{\nu}$ cannot be predicted. It can only be calculated numerically.

\section{Results}
\subsection{Spectra for $\Pm=1$}
\begin{figure}
\begin{center}
\begin{tabular}{@{}c@{\hspace{0em}}c@{\hspace{0em}}c@{\hspace{0em}}c@{}}
   \rotatebox{90}{$\quad \quad \quad \quad \Omega^{-1/2}\varepsilon^{-1/2}k^2E(k)$} & \includegraphics[width=0.45\textwidth]{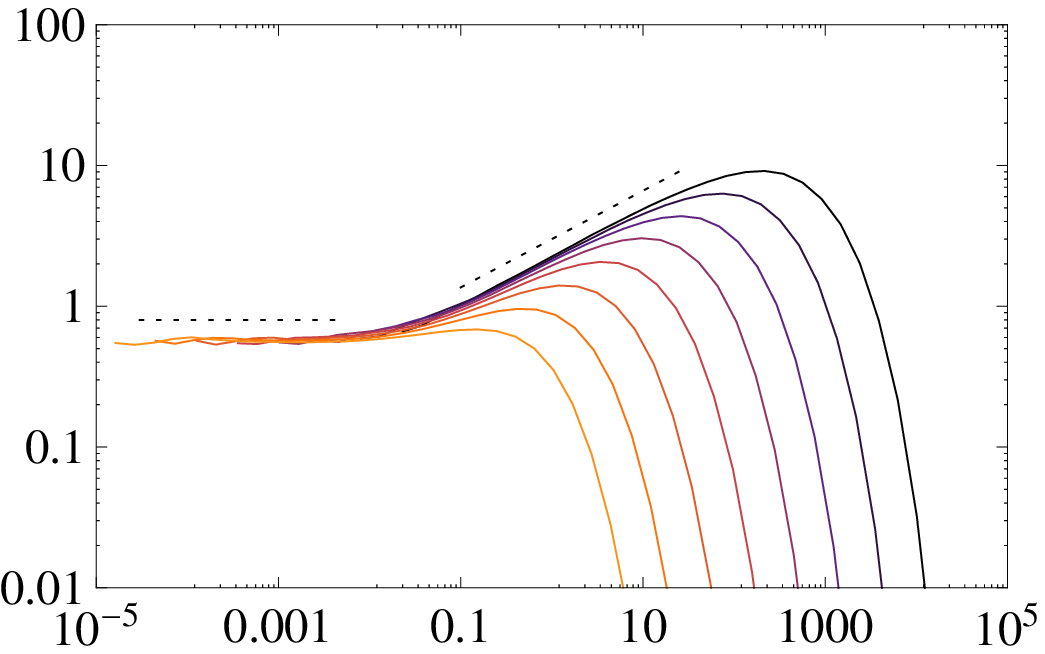}\\															
       & $\Omega^{-3/2}\varepsilon^{1/2}k$ \\
   \rotatebox{90}{\quad \quad \quad \quad $\varepsilon^{-2/3}k^{5/3}E(k)$} & \includegraphics[width=0.45\textwidth]{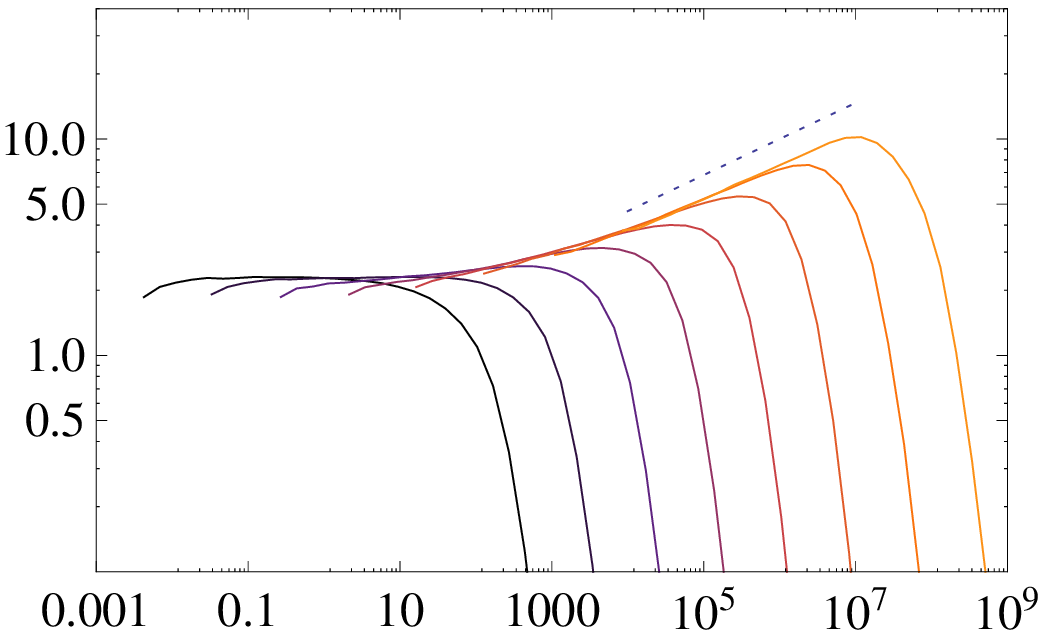}
   \\*[0.0cm]
       & $V_A^{3}\varepsilon^{-1}k$\\
 \rotatebox{90}{\quad \quad \quad \quad $k^{3/2}\varepsilon^{-1/2} V_A^{-1/2}E(k)$} & \includegraphics[width=0.45\textwidth]{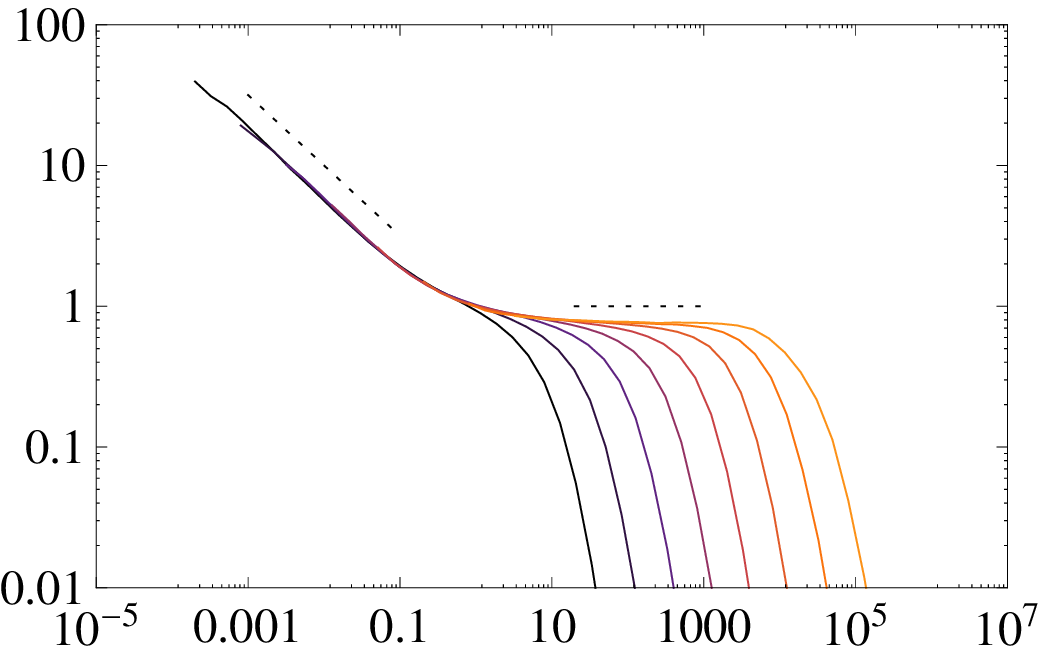}&&\\*[0.0cm]
       & $V_A\Omega^{-1}k$   &&
   \end{tabular}
  \end{center}
\caption{(Color online) Normalized spectra for $\nu=10^{-7}$ and $\Pm=1$. Curves (a) are shown for $V_A=0$ and  $\Omega=12.5,25,50,10,200,400,800,1600$ (from right to left, from darker to lighter).
Curves (b) are shown for $\Omega = 0$ and $V_A=0.16, 0.32, 0.64,1.28,2.56,5.12,10.24,20.48$ (from left to right, from darker to lighter).
Curves (c) are shown for
$(V_A,\Omega)=(0.16,800); (0.32, 400); (0.64,200); (1.28,100); (2.56,50);$ $(5.12,25); (10.24,12.5); (20.48,6.25)$ (from left to right, from darker to lighter).}
\label{Pm=1}
\end{figure}

In Fig.~\ref{Pm=1} the spectra are plotted for $\nu=10^{-7}$ and $\Pm=1$ in the three cases $V_A=0$,
$\Omega = 0$, and $V_A \Omega \ne 0$. For $V_A=0$, the horizontal and $k^{1/3}$ dashed lines
disclose a RK regime. For $\Omega = 0$, the dashed line $k^{1/6}$  disclose a KA regime.
For $V_A \Omega \ne 0$, the $k^{-1/2}$ and horizontal dashed lines disclose a RA regime.
In each case the transition between two power laws is rather smooth and occurs over a scales range
of about two orders of magnitude.

For $\varepsilon \approx 1$ and taking the numerical values for $\Omega, V_A$ and $\eta$
given in Fig.~\ref{map}
we find that
the three sets of spectra found with the shell model belong indeed to the three parts
RK, (R)KA and RA of Fig.~\ref{map}.
We tried to track the transition from one part to the other, varying $\Omega$ and $V_A$. It is however not possible to handle it numerically as the spectral slopes
are not so well defined at the neighborhood of the frontiers delimiting the four parts of Fig.~\ref{map}.

\begin{figure*}
\begin{center}
\begin{tabular}{@{}c@{\hspace{0em}}c@{\hspace{0em}}c@{\hspace{0em}}c@{}}
   \rotatebox{90}{\quad \quad \quad \quad \quad \quad $k^2E(k)\varepsilon^{-1/2}$} &
   \includegraphics[width=0.45\textwidth]{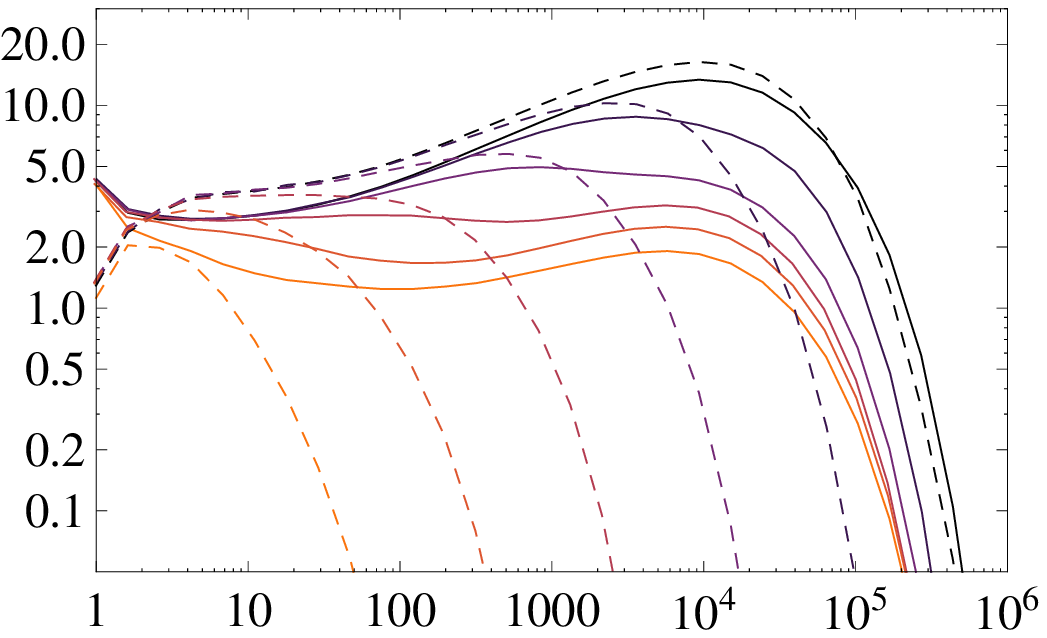}&
   \rotatebox{90}{\quad \quad \quad \quad \quad \quad $k^2E(k)\varepsilon^{-1/2}$}&
   \includegraphics[width=0.45\textwidth]{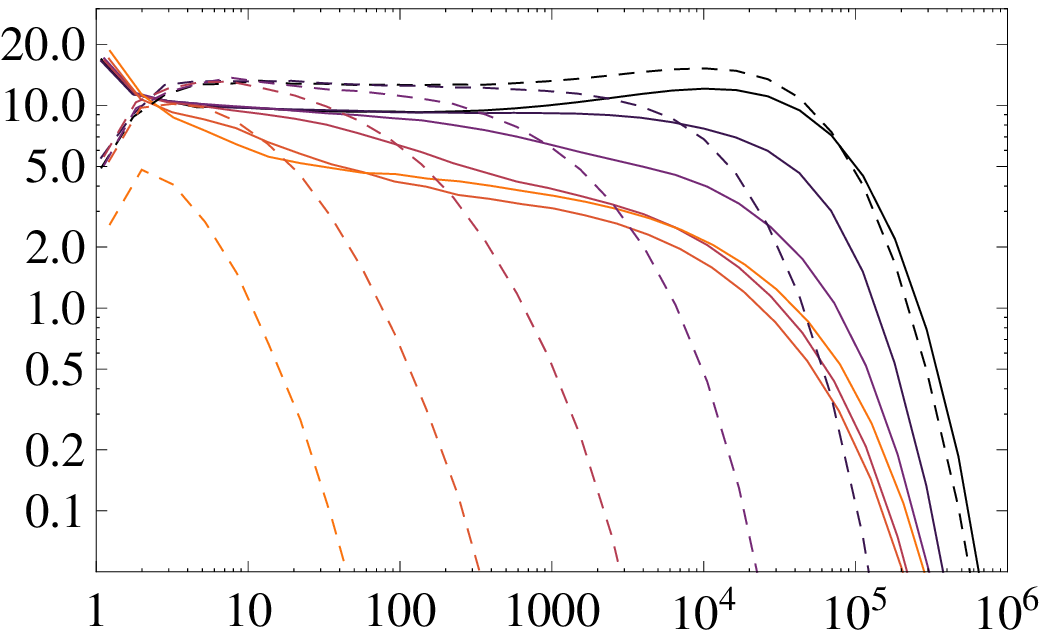}\\*[0.0cm]
       & $k\varepsilon^{1/2}$ && $k\varepsilon^{1/2}$\\*[0.0cm]
          \rotatebox{90}{\quad \quad \quad\quad \quad \quad $k^{5/3}E(k)\varepsilon^{-2/3}$} &
          \includegraphics[width=0.45\textwidth]{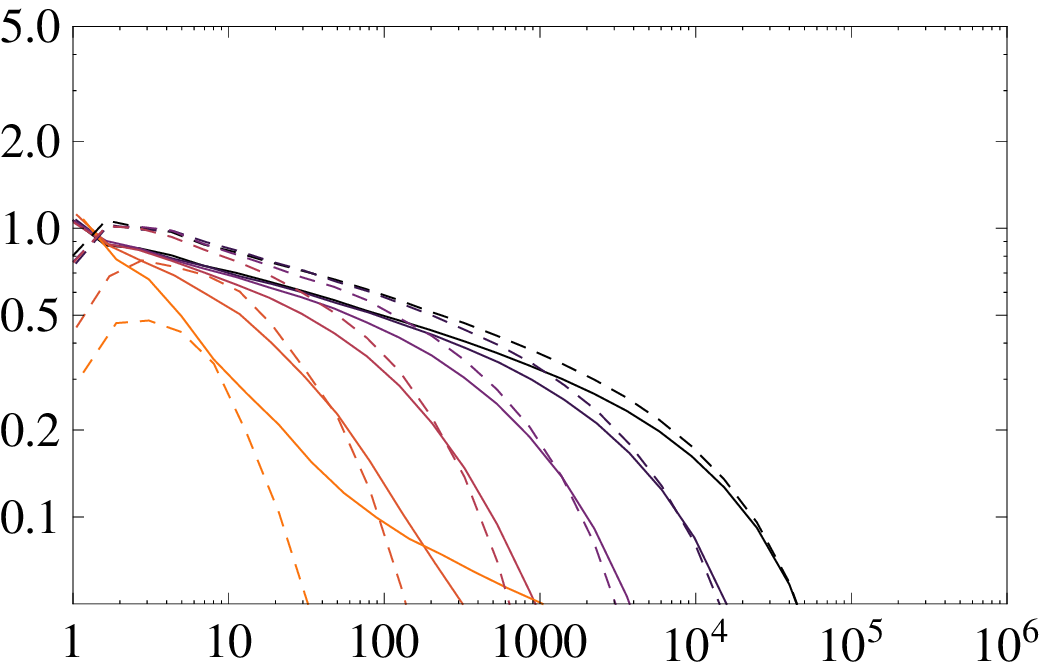}&
          \rotatebox{90}{\quad \quad \quad\quad \quad \quad $k^{3/2}E(k)\varepsilon^{-1/2}$}&
   \includegraphics[width=0.45\textwidth]{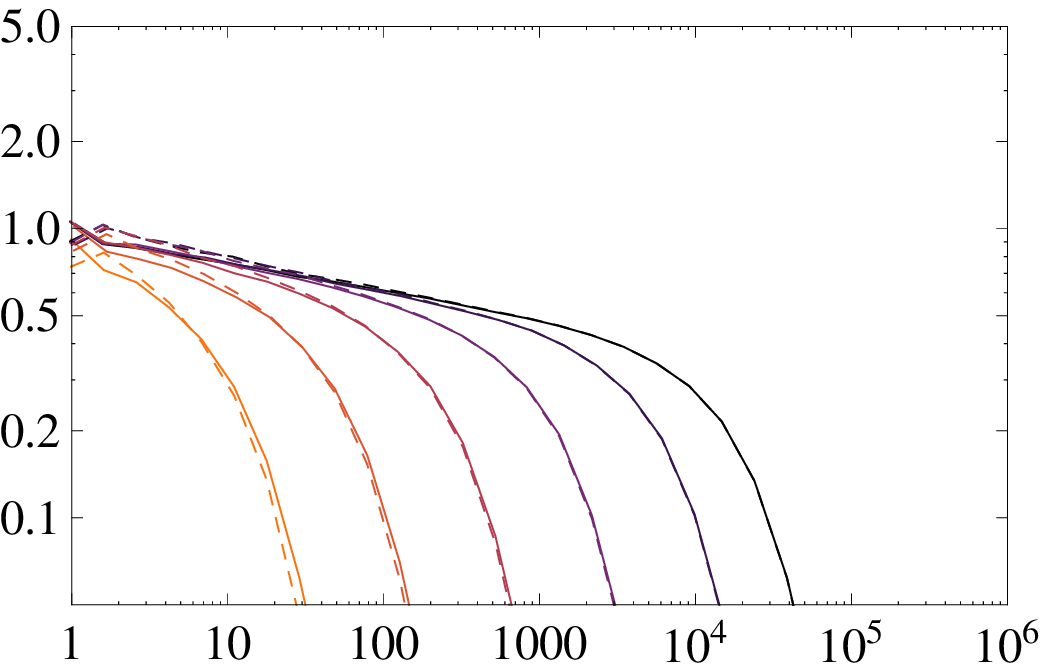}\\*[0.0cm]
       & $k\varepsilon^{-1}$ && $k\varepsilon^{-1}$\\*[0.0cm]
          \rotatebox{90}{\quad \quad \quad\quad \quad \quad $k^{2}E(k)\varepsilon^{-1/2}$} &
   \includegraphics[width=0.45\textwidth]{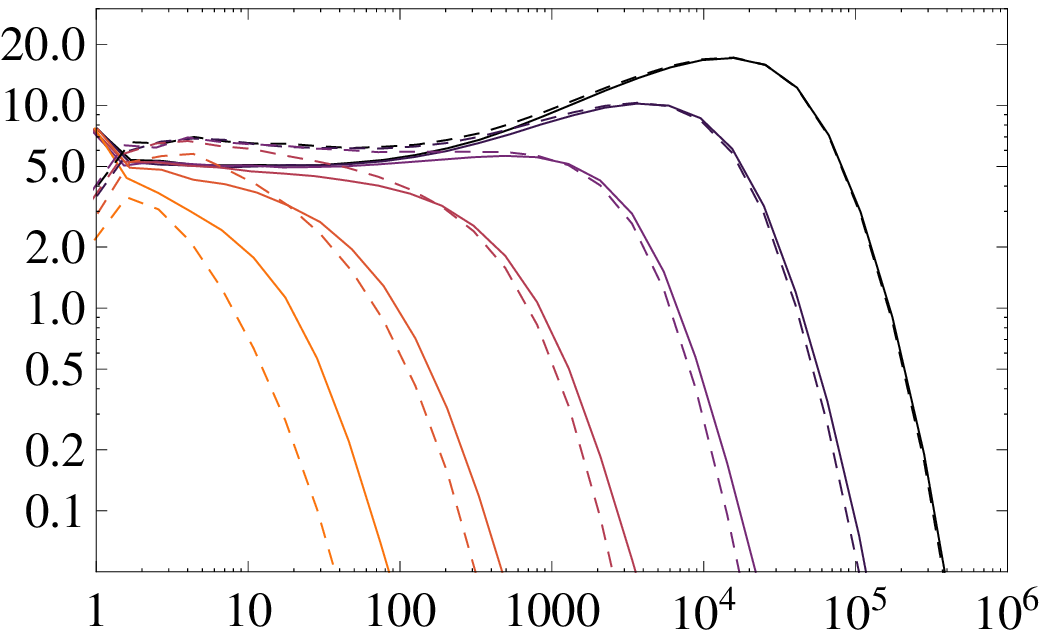}&
   \rotatebox{90}{\quad \quad \quad\quad \quad \quad $k^{3/2}E(k)\varepsilon^{-1/2}$} &
   \includegraphics[width=0.45\textwidth]{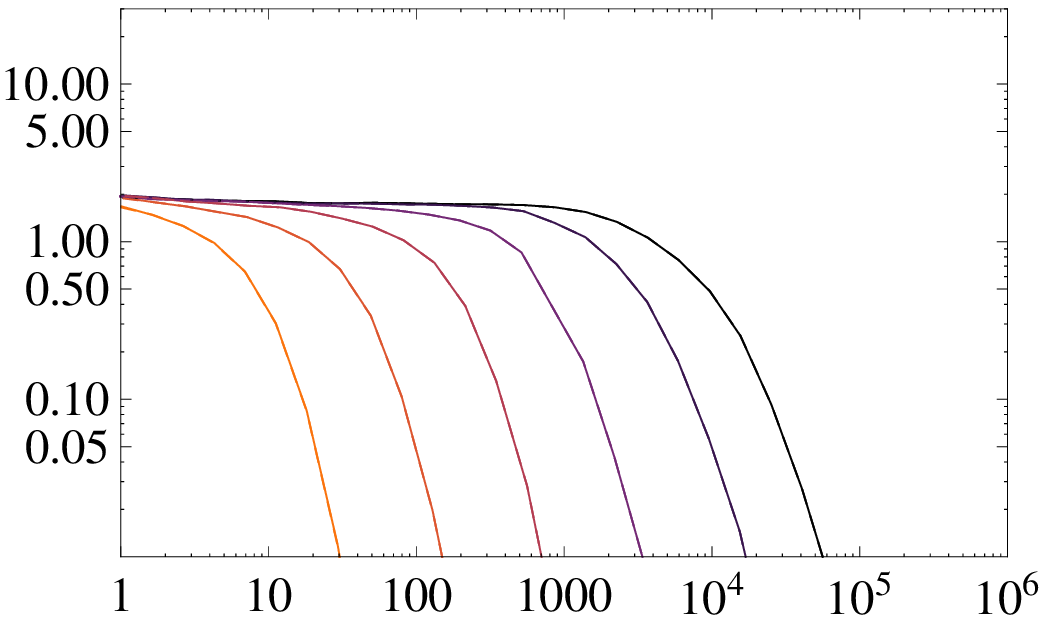}\\*[0.0cm]
       & $k\varepsilon^{-1}$ && $k\varepsilon^{-1}$
   \end{tabular}
  \end{center}
\caption{(Color online) Normalized kinetic (full) and magnetic (dashed) spectra for $\nu=10^{-7}$:
(a) $(V_A,\Omega)=(0,100)$, (b) $(V_A,\Omega)=(0,1600)$.
(c) $(V_A,\Omega)=(0.08,0)$, (d) $(V_A,\Omega)=(1.28,0)$.
(e) $(V_A,\Omega)=(0.32,400)$, (f) $(V_A,\Omega)=(20.48,6.25)$.
For each set of curves for
$\Pm=10^{-5},10^{-4},10^{-3},10^{-2},10^{-1}, 1$ (from lighter to darker). Note, that the kinetic and magnetic spectra are superposed in the case (f).
}
\label{Pm<1}
\end{figure*}

\subsection{Spectra for $\Pm<1$}
In Fig.~\ref{Pm<1} the kinetic and magnetic spectra are plotted for $\nu=10^{-7}$ and several values of $\Pm$,
for the three previous cases.

For $V_A=0$ (a,b) increasing $\Pm$ decreases the magnetic dissipation scale while the viscous scale
is not significantly changed. This is in agreement with a simple Kolmogorov phenomenology
 \cite{Stepanov08}, the ratio of dissipation scales being given by $k_{\nu}/k_{\eta}\propto \Pm^{-3/4}$.
 For $\Pm\ge 10^{-2}$ the effect of rotation is visible in the spectra flatness.
 At smaller values of $\Pm$ it is however difficult to determine any slope at all.

For $\Omega=0$ and $V_A=1.28$ (c,d)
both kinetic and magnetic spectra are almost the same whatever the value of $\Pm$.
The effect of an applied magnetic field is to correlate both fields as expected in Alfv\'en waves.
In particular the dissipation scale is governed by the magnetic diffusivity, with $k_{\nu} \approx k_{\eta}$ .
The same conclusions are found for $(V_A,\Omega)=(0.32,400)$ (e) and $(V_A,\Omega)=(20.48,6.25)$ (f).
In these two cases the horizontal slopes are due to rotation (e) and applied magnetic field (f).

We note that for $\Omega=0$ and $V_A=1.28$ (d) the normalized curves are not horizontal. They correspond
 to spectral energy density slopes between $k^{-5/3}$ and $k^{-3/2}$. The latter is obtained for values of
 $V_A$ about ten times larger.
\begin{figure*}
\begin{center}
\begin{tabular}{@{}c@{\hspace{0em}}c@{}}
\includegraphics[width=0.45\textwidth]{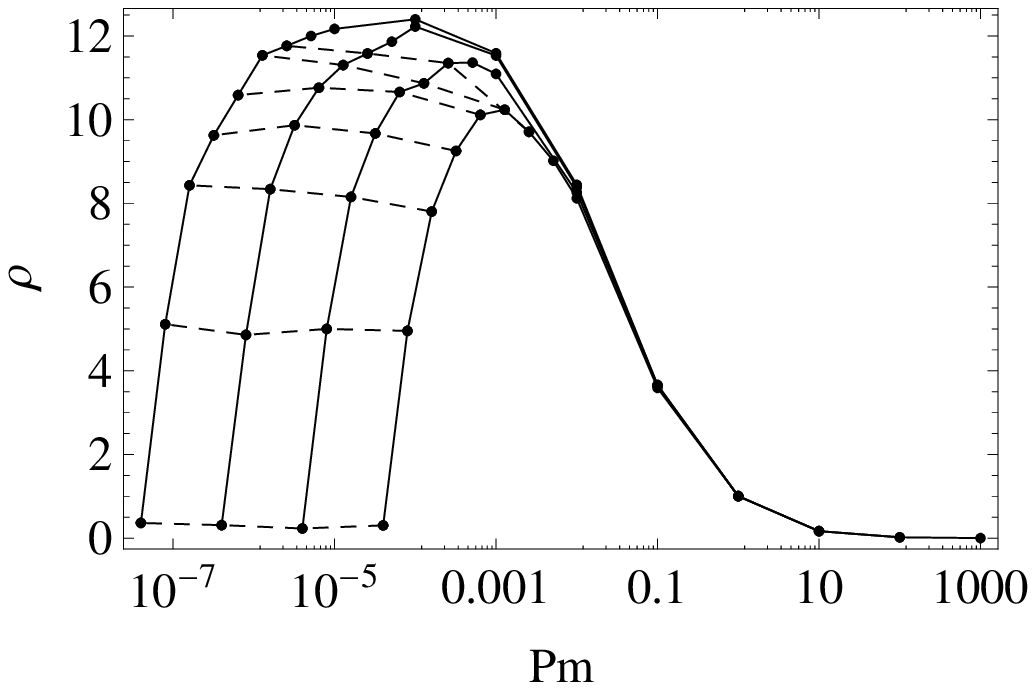}&
\includegraphics[width=0.45\textwidth]{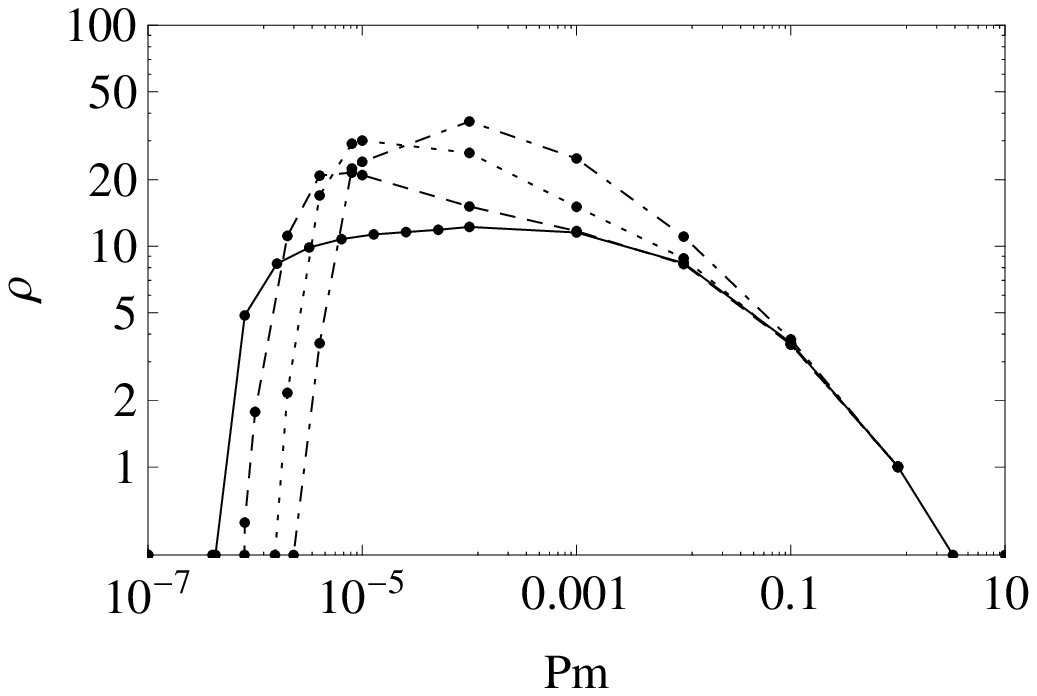}\\
\includegraphics[width=0.45\textwidth]{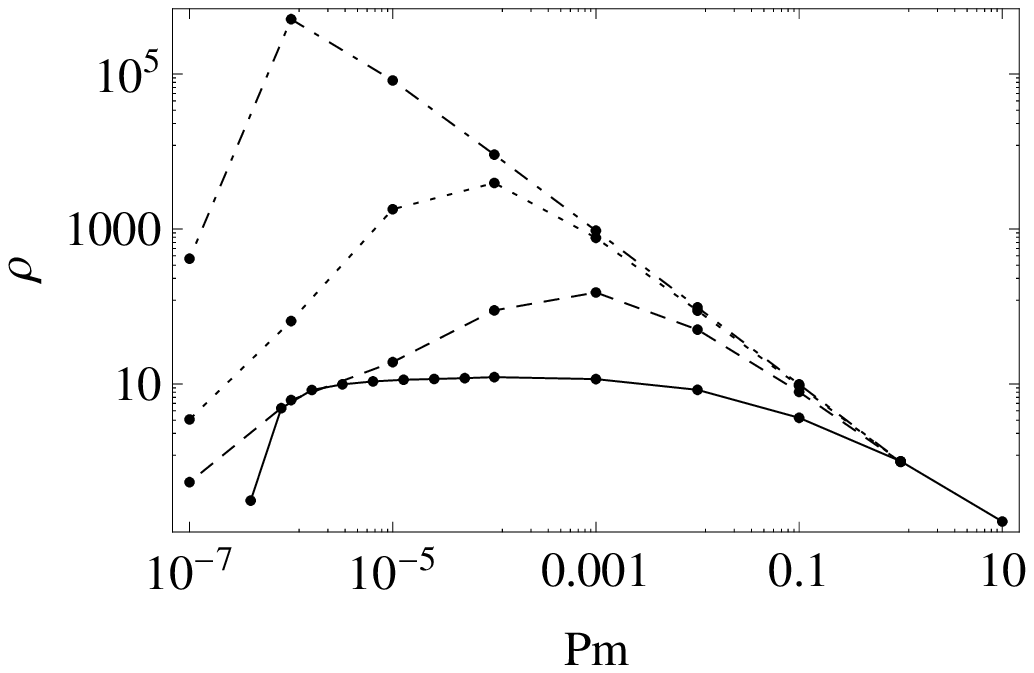}&
\includegraphics[width=0.45\textwidth]{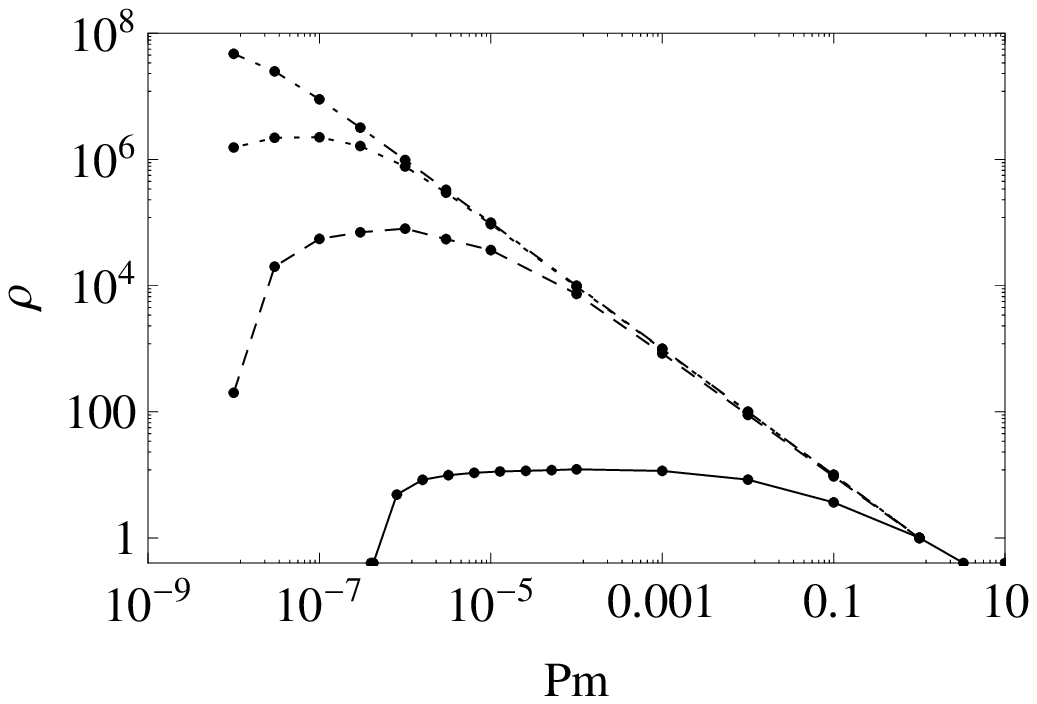}
   \end{tabular}
  \end{center}
\caption{Dissipation ratio versus $\Pm$.
On  panel (a) $V_A=\Omega=0$ and the full curves from right to left correspond to $\nu=10^{-5}$, $10^{-6}$, $10^{-7}$, $10^{-8}$. The dashed curves from bottom to top correspond to $\eta=1/4$, $1/8$, $1/16$, $1/32$, $1/64$, $1/128$, $1/256$.
On  panel (b) $V_A=0$, $\nu=10^{-7}$ and the curves correspond to $\Omega=0$ (full curve), $100$ (dashed), $400$ (dotted), 1600 (dot-dashed).
On  panel (c) $\Omega = 0$, $\nu=10^{-7}$ and the curves correspond to $V_A=0$ (full curve), $0.08$ (dashed), $0.32$ (dotted), 1.28 (dot-dashed).
On  panel (d) $\nu=10^{-7}$ and the curves from bottom to top correspond to
$(V_A,\Omega)=(0,0)$, (0.32, 400), (2.56,50) and (20.48,6.25).
}
\label{dissipation_ratio}
\end{figure*}

\subsection{Dissipation ratio}
In Fig.~\ref{dissipation_ratio} the ratio $\rho=\varepsilon_{\eta}/\varepsilon_{\nu}$ is plotted versus $\Pm$
for $V_A=\Omega=0$ (a), $V_A=0$ (b), $\Omega=0$ (c) and $V_A \Omega \ne 0$ (d).
In the limit $\Pm\rightarrow 0$ the dynamo action does not occur, implying $\rho\rightarrow 0$.
For $\Pm = 1$ both kinetic and magnetic spectra are identical, implying $\varepsilon_{\nu}=\varepsilon_{\eta}=\varepsilon/2$, and then $\rho=1$.
We always find an intermediate value of $\Pm$ for which $\rho$ reaches a maximum.
This is related to a super-equipartition state in which the magnetic energy is higher than the kinetic energy at large scales. Varying $V_A$ and $\Omega$ we find that this maximum value can increase by several orders of magnitude
and that it does not occur at the same $\Pm$.
For the two last cases an asymptotic curve $\rho=O(\Pm^{-1})$ is obtained for large values of $V_A$.
This is a direct consequence of the equipartition regime $|U_n| \approx |B_n|$ obtained at any scale (see Fig.~\ref{Pm<1}).
In that cases the definition of $\rho$ directly implies the scaling $O(\Pm^{-1})$.

\section{Discussion}
For $\Pm=1$ both approaches, phenomenological and shell model, give consistent results in terms of inertia regimes. They are controlled by the shortest time-scale corresponding either to rotation, applied magnetic field, inertia, or a combination of them. For $\Pm <1$ the magnetic dissipation occurs at a scale larger than the viscous scale implying that the different regimes are not so easy to discriminate. However for a sufficiently strong applied magnetic field both kinetic and magnetic energy spectra are merged, implying a strong increase of the viscous dissipation scale. Whether this is due to our isotropic assumption is not clear and cannot be answered with our models. A consequence is that, for a strong applied field, the ratio of magnetic to kinetic dissipation scales like $O(\Pm^{-1})$ and can reach very high values for $\Pm \ll 1$. Without applied field, this ratio is also maximum for some value of $\Pm \ll 1$, depending on the fluid viscosity  and global rotation.

\begin{acknowledgments}
This work benefited from the support of a RFBR/CNRS 07-01-92160 PICS grant and of a Russian Academy of Science  project 09-P-1-1002.
It was also completed during the Summer Program on MHD Turbulence at the Universit\'e
Libre de Bruxelles in July 2009.
We warmly thank G. Sarson for enlightening discussions and anonymous referees for helping us in improvement of the paper.
\end{acknowledgments}

\bibliographystyle{apsrev}
\bibliography{biblio}
\end{document}